\documentclass[12pt]{article}

\usepackage{epsfig}
\usepackage[inkscapelatex=false]{svg}
\usepackage{physics}
\usepackage{geometry}
\usepackage{graphicx}
\usepackage{multirow}
\usepackage{caption}
\usepackage{floatrow}
\usepackage{float}
\usepackage{amsfonts}
\usepackage{amsmath}
\usepackage[%
    colorlinks=true,
    pdfborder={0 0 0},
    linkcolor=red]{hyperref}
\usepackage[english]{babel}
\usepackage{cite}
\usepackage{subcaption} 
\usepackage[export]{adjustbox} 
\usepackage{cleveref} 

\begin{document}
\title{Ultrafast Spectroscopy of Dirac Semimetal Cd$_3$As$_2$ under Pressure}
\author{Vikas Arora$^{1,2}$, D. V. S. Muthu$^{1}$, R Sankar$^{3}$, A K Sood$^{*1,2}$}
\date{\today}
\maketitle

\begin{center}
\textit{$^{1}$Department of Physics, Indian Institute of Science, Bangalore 560012, India\\
$^{2}$Center for Ultrafast Laser Applications, Indian Institute of Science, Bangalore 560012, India\\
$^{3}$Institute of Physics, Academia Sinica - Taipei 11529, Taiwan}
\end{center}

\begin{center}
\textbf{Keywords:} Dirac semimetal, High Pressure, optical pump-optical probe spectroscopy, ultrafast carrier relaxation dynamics. 
\end{center}


\begin{abstract}
Topological properties of a three-dimensional Dirac semimetal Cd$_3$As$_2$, protected by crystal rotation and time-reversal symmetry, can be tuned with the application of pressure. Ultrafast spectroscopy is a unique tool to investigate the character and time evolution of electronic states, emphasizing the signatures of transition. We designed an experimental setup for in-situ pressure-dependent ultrafast optical pump optical probe spectroscopy of Cd$_3$As$_2$ using a symmetric diamond anvil cell. The fast relaxation processes show significant changes across pressure-induced phase transitions at P$_{C_1}$$\sim$ 3 GPa and P$_{C_2}$$\sim$ 9 GPa. A new sub-picosecond time scale relaxation dynamics emerges beyond P$_{C_2}$. Theoretical calculations of differential reflectivity for both interband and intraband processes indicate that the negative (positive) differential reflectivity ($\Delta$R/R) results from the interband (intraband) processes. The pressure-dependent behavior of relaxation dynamics amplitudes beyond P$_{C_1}$ emphasized the necessity of incorporating quadratic band opening in the calculations, explaining the transition of Cd$_3$As$_2$ from a Dirac semimetal to a semiconducting phase. The time evolution of differential reflectivity is calculated using the electronic temperature as a function of time, as provided by the two-temperature model, which fits the experimental data.
%

\end{abstract}


\section*{Introduction}
Topological quantum materials exhibit diverse physical phenomena, including anomalous Hall effect \cite{Wang_2018}, dissipationless transport with giant mobilities \cite{Shekhar_2015, Liang_2015}, magneto-optic Kerr effects \cite{Higo_2018}, large magnetoresistance \cite{Shekhar_2015}, and Nernst effect \cite{Ikhlas_2017} that violates the classical Wiedemann-Franz law \cite{Gooth_2018}. Topology of band structure plays a crucial role in materials such as topological insulators, Dirac and Weyl semimetals. Topological insulators feature insulating bulk states along with the topological metallic Dirac surface states protected by time-reversal symmetry. The locking of electronic spin and momentum in a perpendicular direction due to spin-orbit coupling facilitates low-dissipation transport by suppressing back-scattering \cite{Kumar_2021}. A similar phenomenon is observed in three dimensions for Dirac semimetals (DSM) and Weyl semimetals, where nontrivial topological bulk states are protected by crystal symmetry, such as mirror or rotation. Weyl semimetals can be observed in non-centrosymmetric crystals or systems lacking time-reversal symmetry. Interestingly, the parent material, DSM, can be transformed into Weyl semimetals (WSM) by breaking inversion symmetry or time-reversal symmetry through the application of a magnetic field or circularly polarized light. Additionally, they can be tuned into topological insulators by enhancing spin-orbit coupling. In Dirac semimetals (DSM), the valence and conduction bands meet at Dirac points and disperse linearly in all three directions of momentum, analogous to the behavior observed in two-dimensional graphene \cite{Young_2012,Yang_2014,Liu_2014,Burkov_2011,Wang_2013}. In support of the theoretical predictions\cite{Wang_2013}, the presence of 3D Dirac points is evidenced by scanning tunneling microscopy\cite{Jeon_2014} and angle-resolved photoemission spectroscopy\cite{Yi_2014,Liu_2014, Neupane_2014,Borisenko_2014}. The DSM with non-trivial topology shows very high mobility\cite{Rosenberg_1959}, giant magnetoresistance and diamagnetism\cite{Liang_2015,Li_2015,Narayanan_2015}, and low dissipation Dirac transport\cite{Wang_2013,Liu_2014}. Graphene has applicability in photonics such as optical modulators\cite{Sun2016}, mid-infrared photodetectors\cite{Xia2009}, and tunable ultrafast laser\cite{Martinez2013}, but the poor absorption of photons limits the applicability of these devices. However, with an extra dimension, DSM as a three-dimensional analog of graphene can be a better candidate\cite{Yao_2021}. 
\par
Examples of DSM are A$_3$Bi (A=K, Rb, Na)\cite{Wang_2012,Liu2014}, $\beta$-cristobalite BiO$_2$\cite{Young_2012} and Cd$_3$As$_2$\cite{Neupane_2014,Wang_2013}. The first two are air-sensitive and unstable at ambient conditions whereas Cd$_3$As$_2$ is stable. The ultra-high mobility of $\sim$10 m$^2$V$^{-1}$s$^{-1}$ for Cd$_3$As$_2$ at room temperature has been reported\cite{Rosenberg_1959}. Cd$_3$As$_2$-based photodetectors for the spectral range of 532 nm - 10.6 $\mu$m have been demonstrated with a high bandwidth of 145 GHz, which provides a broad wavelength range\cite{Wang2017}. For the mid-infrared range (3-6 $\mu$m), Cd$_3$As$_2$ has shown excellent saturable absorption properties and they can be used as passive mode-lockers to generate ultrashort pulses\cite{Meng_2018}. The possible applications of DSM Cd$_3$As$_2$ in electronic and optoelectronic devices require the understanding of transient dynamics of the carriers, the dynamical processes involved between carriers and phonons as well as other quasiparticles to better understand the device performance. A two-step cooling mechanism of hot carriers has been presented in the THz studies and the transient reflectance of Cd$_3$As$_2$\cite{WeiLu_2018,Weber_2015,WeiLu_2017}. The electron-phonon coupling factor $g$=5.3$\times$10$^{15}$ Wm$^{-3}$K$^{-1}$ has been reported by broadband hot carrier dynamics for Cd$_3$As$_2$\cite{Zhu2017}, which is similar to graphite but smaller than noble metals. With the help of Cr doping of Cd$_3$As$_2$ (replacing a few Cd sites), the relaxation time of photoexcited carriers can be controlled from 8 ps to 800 fs at a probe wavelength of 4.5 $\mu$m, lending support to the development of mid IR ultrafast sources with high performance\cite{Zhu_2017}. The symmetry breaking of DSM Cd$_3$As$_2$ can be achieved by chemical doping or the application of pressure. The former method brings chemical complexity whereas the pressure is a cleaner tool. 
An insulator-to-metal transition is observed in Pb$_{1-x}$Sn$_x$Te at 1 GPa, exhibiting a metallic state with large Berry curvature and negative magnetoresistance, indicative of a Weyl semimetal (WSM) \cite{Liang_2017}. \textit{T$_d$}-MoTe$_2$ transitions to a topological superconductor below a temperature of 0.1 K can be achieved at 8 K under 12 GPa pressure\cite{Qi_2016}. Superconductivity in \textit{T$_d$}-MoTe$_2$ at ambient pressure occurs in the orthorhombic phase, extending to a coexistent phase of orthorhombic (\textit{T$_d$}) and monoclinic (1\textit{T'}) structures under pressure \cite{Dissanayake_2019}. The Fermi surface topology in WSMs can be probed by pressure dependence of quantum oscillations\cite{Reis_2016}. Near-infrared and mid-infrared spectroscopy reveal the pressure-induced insulator-to-metal transition in VO$_2$, showing the electronic character while preserving the monoclinic structure\cite{Braun_2018}. A few recent articles provide an overview of ultrafast dynamics under high pressure for materials such as black phosphorus \cite{Wu_2024}, zirconium tritelluride \cite{Zhang_2024}, as well as two-dimensional materials, quantum dots, and perovskites \cite{Tu_2023}. The pressure-driven suppression of a spin density wave in iron pnictide BaFe$_2$As$_2$ is shown to exhibit the second-order character of a quantum phase transition, in contrast to the first-order temperature-induced SDW transition\cite{Fotev_2023}. A recent report on the topological insulator Bi$_2$Se$_3$ indicates an electronic topological transition through an anomaly in relaxation time around 3 GPa\cite{Yang_2024}. 
The electrical transport studies and synchrotron X-ray diffraction studies of Cd$_3$As$_2$ have revealed a phase transition from metallic tetragonal phase to semiconducting monoclinic phase at a pressure of 2.6 GPa \cite{Zhang2015}. Further, the activation energy of the semiconductor phase shows a jump after 9 GPa \cite{Zhang2015}. The high-pressure transition was confirmed by Raman spectroscopy\cite{Gupta_2017} at $\sim$ 9.5 GPa. The carrier relaxation dynamics in Cd$_3$As$_2$ has been limited to the ambient pressure studies. The versatile characteristics of this material motivated us to explore the carrier relaxation dynamics under pressure.
\par
Here, we report the pressure-dependent optical pump-optical probe spectroscopy of Cd$_3$As$_2$ with an in-situ setup using the diamond anvil cell (DAC), shown in Figure \ref{Fig:3p1}. The differential reflectivity ($\Delta$R/R) initially shows a negative behavior, followed by a positive value at longer times. The amplitude and the relaxation time of the fast decay process show significant changes across the transition pressures, P$_{C_1}$ $\sim$ 3 GPa and P$_{C_2}$ $\sim$ 9 GPa. Beyond the pressure P$_{C_2}$, we observe the unfolding of a new and very fast ($<$1 ps) decay process. Theoretical calculations for differential reflectivity are reported, incorporating interband and intraband contributions and a progressive band opening beyond P$_{C_1}$, closely understanding our experimental results. The two-temperature model (TTM) combined with reflectivity calculations illustrates the time evolution of the differential reflectivity.


\section*{Experimental details}
Optical pump-optical probe spectroscopy (OPOP) experiments were performed using Ti:Sapphire pulsed laser (M/s Newport Corporation Pvt Ltd) with a pulse width of $\sim$50 fs and a repetition rate of 1 kHz. The pump and probe beams, both having a central photon energy of 1.55 eV, were kept in a cross-polarized configuration. Pressure-dependent OPOP experiments were carried out using a symmetric diamond anvil cell. A pre-indented stainless steel gasket with a hole of diameter $\sim$150 $\mu$m was placed in between the diamond anvils to act as a sample chamber. A small piece of crystalline Cd$_3$As$_2$ (size $\sim$70 $\mu$m) was kept along with a ruby chip (pressure calibration) and NaCl (serving as pressure transmitting medium) in the sample chamber. The ruby fluorescence technique was used to calibrate the pressure inside the sample chamber\cite{Mao1978, Mao1986, Prut2022}. The ultrafast measurements were carried out from the front side of the DAC, whereas the ruby fluorescence for the pressure calibration was recorded from the back side of the DAC. The ruby fluorescence was performed using a cw He-Ne laser of wavelength 632 nm coupled to a spectrometer (iHR320 by M/s Horiba Jobin Yvon). Single crystal of Cd$_3$As$_2$ was grown using the self-selecting vapour growth (SSVG) method. The crystals obtained are n-doped. The samples were characterized using scanning tunneling microscopy, transmission electron microscopy, transport property measurements, angle-resolved photoemission spectroscopy\cite{Sankar_2015} and Raman spectroscopy \cite{Gupta_2017}.
\par
\begin{figure}[H]
   \centering
   \includegraphics[width=\textwidth]{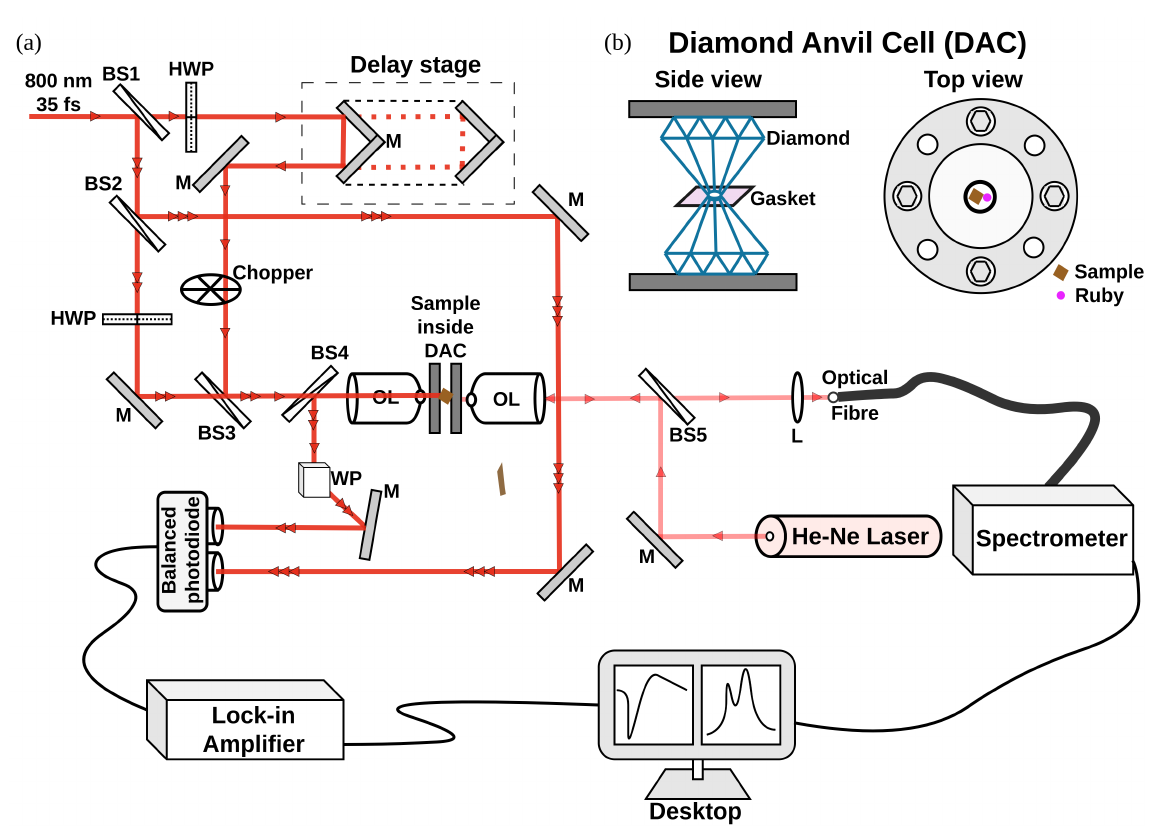}
    \caption[In-situ high-pressure experimental setup for ultrafast optical pump-optical probe spectroscopy]{(a) In-situ high-pressure experimental setup for ultrafast optical pump optical probe-spectroscopy. The left half section (with dark red beam) represents the ultrafast spectroscopy measurements, whereas the right section (bottom half, with faint red beam) represents the pressure calibration setup for recording the fluorescence of Ruby, (b) The zoomed-out version of `Sample inside DAC' where Diamond Anvil Cell (DAC) with sample loaded (side and top view) is depicted. BS: Beam Splitter, HWP: Half Wave Plate, M: Mirror, WP: Wollaston Prism, OL: Objective Lens, L: Lens.}
    \label{Fig:3p1}
\end{figure}

Figure \ref{Fig:3p1}\textcolor{red}{a} shows the experimental setup for the in-situ pressure-dependent optical pump optical probe (OPOP) spectroscopy. The left half section with the bold red beam path represents the OPOP measurements. The pump beam (single arrow) and the probe beam (double arrow) are made collinear to pass through an objective lens (OL) to fall on the sample inside the DAC. The reflected beams retrace the path through OL to the beam splitter (BS4) and are made to pass through Wollaston prism (WP). The cross-polarized pump and probe beams separate from each other using the WP and the latter one is sent to a balanced photodetector, where a reference beam (triple arrow) is also being collected. The photoresponse obtained is amplified by a lock-in amplifier and the data are recorded using the LabVIEW interface. A CCD camera, positioned before the WP and utilizing an additional beam splitter (not depicted here) is employed to image the sample chamber. This camera is oriented perpendicularly to the plane of beam propagation.
\par
The right bottom section of the setup (Figure \ref{Fig:3p1}\textcolor{red}{a}) represents the recording of Ruby fluorescence. A cw laser beam of wavelength 632 nm (faint red color, single arrow) is made to pass through the objective lens (OL) to fall on the ruby. The scattered light from the ruby is collected through a beam splitter and a biconvex lens (L) onto the optical fiber. The optical fiber sends the scattered light to a spectrometer and the fluorescence spectrum of ruby is recorded. A side and top view of a symmetric diamond anvil cell (DAC) is presented in the top right section of Figure \ref{Fig:3p1}\textcolor{red}{b}. The two screws of the symmetric DAC are rotated clockwise and the other two in counter-clockwise directions (arranged alternatively) to tune the pressure. This process is performed in-situ without taking DAC out of the beam path to carry out OPOP experiments at different pressures up to 11.3 GPa.


\section*{Results and discussions}


\section{Experimental Results}

\begin{figure}[H]
    \centering
    \includegraphics[width=\textwidth]{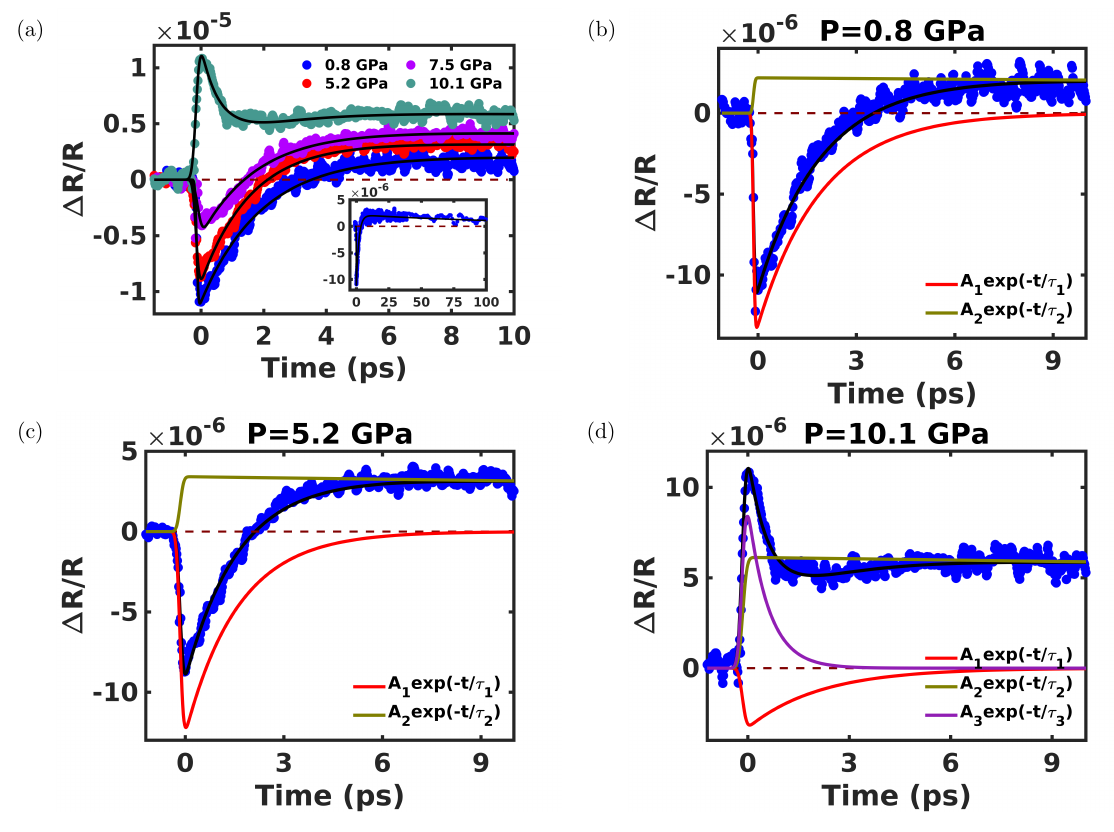}
    
     \caption[Differential reflectivity $\Delta$R/R of DSM Cd$_3$As$_2$ at various pressures]{(a) $\Delta$R/R with real-time: The solid dots represent the data at various pressures, while the black solid lines denote the decay processes provided by Eqs.\ref{eq:3p1}-\ref{eq:3p2}. The cumulative plot of $\Delta$R/R at different pressures indicates the tunability of relaxation dynamics when subjected to external pressure. The inset shows the biexponential fitting up to 100 ps at P=0.8 GPa.(b) and (c) The slow (olive green) and fast (red) components of the fitting model are presented for the pressures of 0.8 GPa and 5.2 GPa, respectively. (d) An additional fastest component (purple) is necessary to fit the data above a pressure of 9 GPa (Eq.\ref{eq:3p2}).}
    \label{Fig:3p2}
\end{figure}

Figure \ref{Fig:3p2}\textcolor{red}{a} shows the time evolution of differential reflectivity, $\Delta$R/R at pressures of 0.8 GPa, 5.2 GPa, 7.5 GPa, and 10.1 GPa using the pump fluence of 243 $\mu$J/cm$^2$. The data (solid dots) is well fitted by a biexponential function for pressures below 9 GPa, as indicated by the black curve. The inset shows the biexponential model fitting up to 100 ps.
 \begin{equation}
    \centering
    \frac{\Delta R}{R}(t)=\frac{1}{2}\large(1+erf\large(\frac{t-t_0}{\tau_{r}}\large)\large)(A_1e^{-t/\tau_1}+A_2e^{-t/\tau_2})
    \label{eq:3p1}
\end{equation}
where (A$_1$, $\tau_1$) and (A$_2$, $\tau_2$) represent the fast (negative component) and slow (positive component) relaxation processes respectively. The fast and slow components are represented by red and olive curves in Figure \ref{Fig:3p2}\textcolor{red}{b} and Figure \ref{Fig:3p2}\textcolor{red}{c}, corresponding to the pressures of 0.8 GPa and 5.2 GPa. Following the pump excitation, the hot carriers attain a quasi-thermal equilibrium with high electronic temperature, T$_e$, within tens of fs due to e-e collisions\cite{Takeda_2024}. Afterwards, they start cooling by the emission of hot optical phonons which further cool by giving their energy to the acoustic phonons. This is the two-temperature model of the carrier relaxation process \cite{chen2006}. To investigate the behavior of these relaxation processes under high pressure, we have recorded the $\Delta$R/R at various pressures. Figure \ref{Fig:3p2}\textcolor{red}{a} displays the time evolution of $\Delta$R/R at four different pressures. With an increase in pressure beyond P=3 GPa, the maximum of $\Delta$R/R starts to decrease and turns positive beyond 9 GPa, demonstrating the pressure-induced control of relaxation dynamics in Cd$_3$As$_2$. The experimental time data of $\Delta$R/R can be fitted by Eq. \ref{eq:3p1} until $\sim$ 9 GPa. Beyond 9 GPa, the $\Delta$R/R shows a different trend, namely positive $\Delta$R/R over all the time scales. This data can only be fitted by Eq.\ref{eq:3p2}, as illustrated in Figure \ref{Fig:3p2}\textcolor{red}{d}, which necessitated an additional faster component.
 \begin{equation}
    \centering
    \frac{\Delta R}{R}(t)=\frac{1}{2}\large(1+erf\large(\frac{t-t_0}{\tau_{r}}\large)\large)(A_1e^{-t/\tau_1}+A_2e^{-t/\tau_2}+A_3e^{-t/\tau_3})
    \label{eq:3p2}
\end{equation}
where the amplitude A$_3$ is positive and the relaxation time $\tau_3$ is $\sim$0.7 ps.
\par
\begin{figure}[H]
    \centering
    \includegraphics[width=\textwidth]{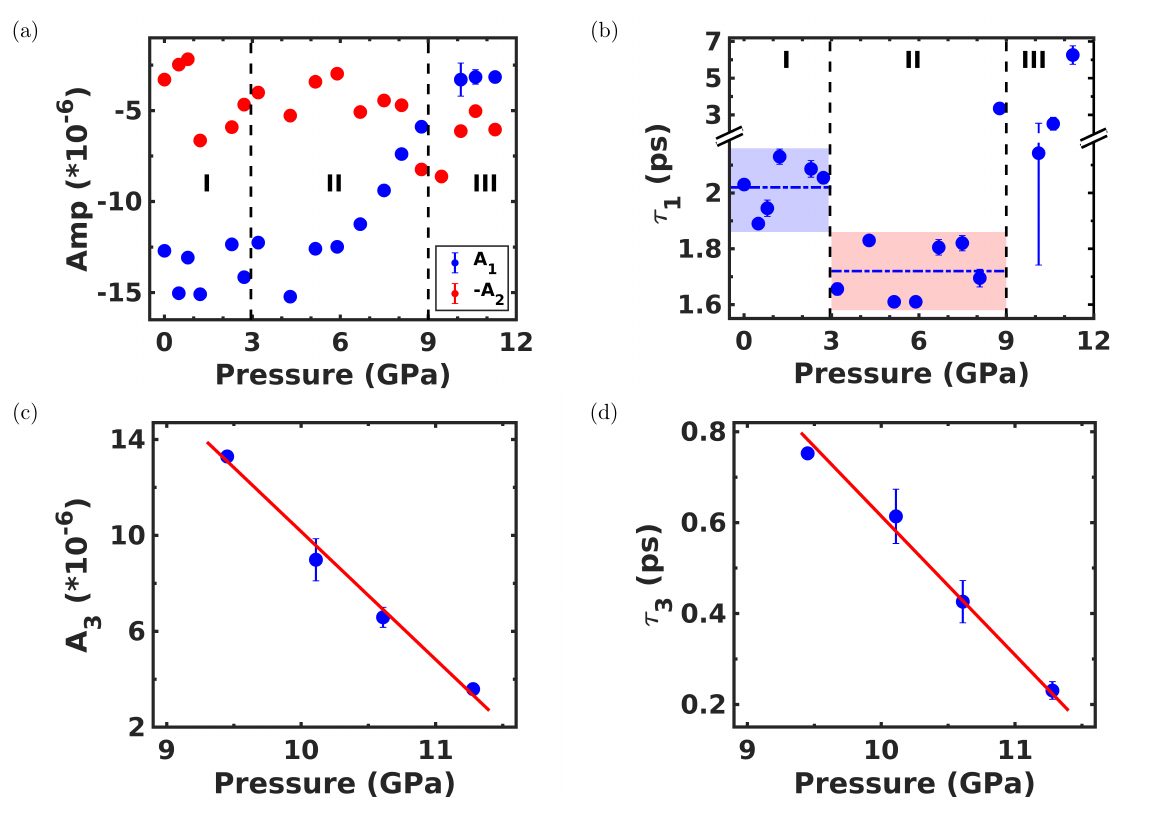}
    \caption[Amplitudes and relaxation times of carrier dynamics in DSM Cd$_3$As$_2$ as a function of pressure]{(a) The fast relaxation amplitude (A$_1$) remains constant in the region I, gradually decreases in region II and remains constant again in region III. The amplitude A$_2$ is weak as compared to A$_1$ in regions I and II, (b) The fast relaxation time ($\tau_1$) is 2.0$\pm$0.1 ps in region I, shifts to 1.7$\pm$0.1 ps in region II and exhibits a drastic change in region III, For P=9.5 GPa, A$_1$ is close to zero and hence it is not shown in (a) and (b). (c) The fastest relaxation amplitude (A$_3$) and (d) its relaxation time ($\tau_3$), which become evident in region III only, decrease with increase in pressure.
}
    \label{Fig:3p3}
\end{figure}

Figure \ref{Fig:3p3}\textcolor{red}{a} represent the amplitudes A$_1$ and A$_2$ as a function of pressure, while Figure \ref{Fig:3p3}\textcolor{red}{b} depicts the pressure dependence of the fast relaxation time, $\tau_1$. The amplitude A$_1$ remains relatively constant in region I (0$<$P$\le$3 GPa), decreases gradually in region II (3$<$P$<$9 GPa) and constant in region III (P$>$9 GPa), as shown in Figure \ref{Fig:3p3}\textcolor{red}{a}. Figure \ref{Fig:3p3}\textcolor{red}{b} shows that the relaxation time $\tau_1$ is 2.0$\pm$0.1 ps in region I, showing a decrease to 1.7$\pm$0.1 ps in region II. The relaxation time $\tau_1$ is associated with the electron-phonon scattering process, where thermally equilibrated photoexcited electrons cool down through the emission of optical phonons\cite{WeiLu_2017,WeiLu_2018}. Beyond 3 GPa, the jump in $\tau_1$ can be understood as a change in electron-phonon coupling, suggesting a phase transition of Cd$_3$As$_2$ across P$_{C_1}$\cite{Zhang2015}. It has been reported that DSM Cd$_3$As$_2$ shows a transition from Dirac semimetal to semiconductor across the pressure of P$\sim$2.5 GPa\cite{Gupta_2017,Zhang2015}. The slow decay process, characterized by amplitude A$_2$ and the relaxation time $\tau_2$ (Eq.\ref{eq:3p1}), shows no systematic behavior with pressure. The relaxation time $\tau_2$ fluctuates in the range of 150$\pm$50 ps. Beyond the pressure of P=9 GPa, $\tau_1$ shows an increase.
\par
A new fast relaxation channel represented by (A$_3$,$\tau_3$) in Eq. \ref{eq:3p2} is required for P$>$9 GPa. These parameters (A$_3$, $\tau_3$) where A$_3$ is positive are shown in Figure \ref{Fig:3p3}\textcolor{red}{c}, \ref{Fig:3p3}\textcolor{red}{d}, which clearly reveal the decrease of A$_3$ and $\tau_3$ with the increasing pressure. This transition at P=9 GPa is marked by an additional intraband relaxation for the photoexcited carriers and coincides with the recent report of an iso-structural transition at $\sim$9 GPa\cite{Gupta_2017}. Across this second transition, the activation energy of semiconducting Cd$_3$As$_2$ jumps from $\sim$200 meV to $\sim$400 meV (Figure \ref{Fig:3p4}\textcolor{red}{b})\cite{Zhang2015}. The origin of the new sub-picosecond relaxation dynamics beyond P = 9 GPa remains unclear. However, we can elucidate the origin of the first two relaxation processes through the aid of differential reflectivity calculations.


\section{Theoretical Insights}

We will now understand the physical origin of the negative A$_1$ and positive A$_2$ and A$_3$ components of differential reflectivity, $\Delta$R/R. The negative component of differential reflectivity can be attributed to Pauli blocking or band-filing, where the pump beam depletes the ground state, leaving limited states available for probe absorption due to Pauli's exclusion principle, thus reducing reflectivity\cite{Weber_2015,Wang2019}. The differential reflectivity due to intraband relaxation process will be shown to result in positive $\Delta$R/R. Before we explore the time evolution of $\Delta$R/R, we will first analyze the pressure-dependent behavior of amplitudes A$_1$ and A$_2$, where their sum, A$_1$+A$_2$, represents the differential reflectivity at zero delay time. 
\par
Starting from the reflection coefficient for an electromagnetic wave traveling at normal incidence from air to medium of refractive index $\Tilde{n}=n+i\kappa$, the reflectivity is given by
\begin{equation}
    \centering
    R=\bigg|\frac{1-\Tilde{n}}{1+\Tilde{n}}\bigg|^2
    \label{eq:3p3}
\end{equation}
The differential reflectivity, $\Delta$R/R is given by,
\begin{equation}
    \centering
    \frac{\Delta R}{R} = \frac{1}{R}\left[\frac{\partial R}{\partial n}\Delta n+\frac{\partial R}{\partial\kappa}\Delta\kappa\right]=\frac{1}{R}\left[\frac{4(n^2-\kappa^2-1)}{((n+1)^2+\kappa^2)^2}\Delta n+\frac{8n\kappa}{((n+1)^2+\kappa^2)^2}\Delta\kappa\right]
    \label{eq:3p4}
\end{equation}
where $\Delta n$\ =\ $n(T_e)$\ -\ $n(300 K)$ and $\Delta \kappa$\ =\ $\kappa(T_e)$\ -\ $\kappa(300 K)$. The refractive index $n$ and the absorption coefficient $\kappa$ are related to dielectric constant
\begin{equation}
    \centering
    \epsilon(\omega,T_e)=(n+i\kappa)^2=n^2-\kappa^2+i2n\kappa
    \label{eq:3p5}
\end{equation} 
The temperature dependence of $n$ and $\kappa$ are, in turn, related to the real and imaginary parts of $\epsilon(\omega,T_e)$. In our present experiments, $\hbar\omega$=1.55 eV. First, we calculate the maximum value of $\Delta$R/R from different relaxation processes, corresponding to the maximum value of electronic temperature T$_e$ attained by quasiparticles when quasi-thermal equilibrium is reached after electron-electron scattering in a few of tens of fs \cite{Takeda_2024}. There are two relaxation processes: Intraband and interband processes. We will now proceed to calculate the changes in $\Delta n(T_e)$ and $\Delta\kappa(T_e)$ from the real and imaginary parts of the dielectric function $\epsilon(\omega,T_e)$.
\par
\begin{figure}[H]
    \centering
    \includegraphics[width=\textwidth]{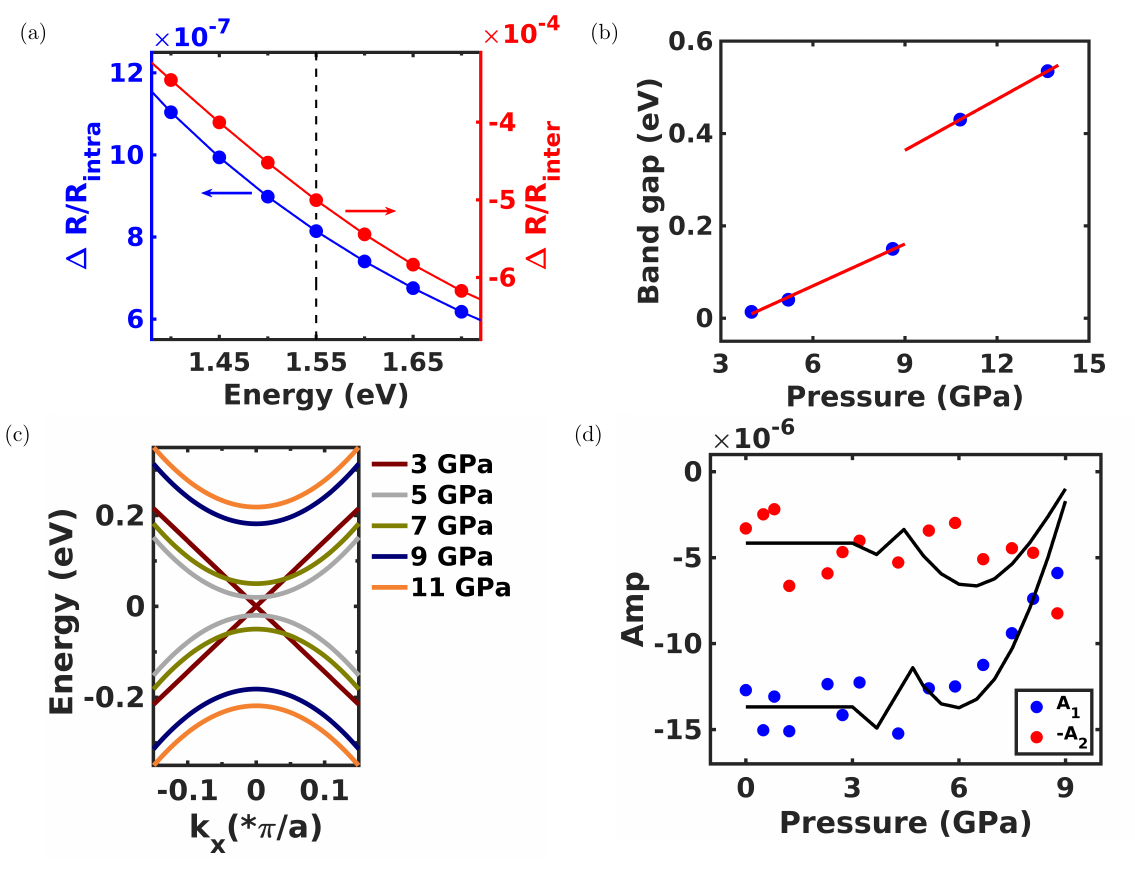}
    \caption[Theoretical calculations for interband and intraband contributions to $\Delta$R/R with pressure]{(a) The differential reflectivity, as a function of energy, is negative for interband (red) and positive for intraband (blue) calculations at 1.55 eV excitation. (b) band gap (the excitation energy) at different pressures, taken from Zhang \textit{et al}\cite{Zhang2015}. The blue circle represents the experimental value, whereas the red curve represents the linear fit for the two different regions, presented by $\eta'$ in the text, (c) the energy eigenvalues of Hamiltonian H (P=3 GPa) and H$_1$ (P$>$3 GPa) or the E-k dispersion relation for various pressures, illustrate the opening of band gap as the pressure is increased beyond 3 GPa, (d) The blue (red) solid dots represent the A$_1$ (A$_2$) with pressure, while the black curves depict the interband (A$_1$) and intraband (A$_2$) components as a function of pressure, based on the calculations described in the text.}
    \label{Fig:3p4}
\end{figure}

The optical absorption will be predominantly influenced by the diagonal elements of the imaginary part of $\epsilon_{ij}(\omega)$, given by \cite{Conte2017}.
\begin{equation}
    \centering
    \epsilon_{jj}(\omega,T_e)=\frac{-4\pi^2e^2}{m^2\omega^2}\frac{1}{V}\sum_k\sum_{c,v}[f(E_{c,k})-f(E_{v,k})]|\bra{ck}p_j\ket{vk}|^2\delta(E_{c,k}-E_{v,k}-\hbar\omega)
\end{equation}
where $V$ is crystal volume, $f(E)$ is the Fermi function and $p$ is the electron momentum operator. We consider interband transitions between occupied Bloch states ($\ket{vk}$) and unoccupied Bloch states ($\ket{ck}$) with $k$ spanning in the Brillouin zone (BZ) with the corresponding energy eigenvalues of E$_{v,k}$ and E$_{c,k}$. The occupancy of these states is characterized by smeared Fermi function $f(E)$. The optical transition strength is governed by the expectation value of the momentum operator $p$. \\
We considered a Dirac Hamiltonian for DSM Cd$_3$As$_2$, $H=\hbar v_F\Vec{k}.\Vec{\sigma}$\cite{Neto_2009} with the energy eigenvalues $\lambda_\pm$=$\pm \hbar v_Fk$=$E_{\pm}$ and the eigenstates $\bra{\lambda_\pm}$=$(1/\sqrt{2k(k\pm k_z)})[k_z\pm k,\ \ k_x+ik_y]$, 
we can rewrite,
\begin{equation}
    \centering
    \epsilon_{jj}(\omega,T_e
    )=\frac{+4\pi e^2}{m^2\omega^2}\frac{1}{V}\sum_k[(f(E)-f(-E))\left[|\bra{\lambda_+}p_j\ket{\lambda_-}|^2\frac{1}{\hbar\omega+i\Gamma-2E}-|\bra{\lambda_-}p_j\ket{\lambda_+}|^2\frac{1}{\hbar\omega+i\Gamma+2E}\right]
    \label{eq:3p6}
\end{equation}
Using Eqs.\ref{eq:3p3}-\ref{eq:3p6}, $\Delta$R/R was calculated for T$_e$=652 K. The electronic temperature, T$_e$ has been evaluated using the two-temperature model, which is described later. It can be seen that the calculations for interband transitions, where $\bra{\lambda_-}$ and $\bra{\lambda_+}$ correspond to the valence band and conduction band, respectively, yield negative $\Delta$R/R (red dots in Figure \ref{Fig:3p4}\textcolor{red}{a}). On the other hand, the intraband calculation for $\Delta$R/R, represented by the blue dots in Figure \ref{Fig:3p4}\textcolor{red}{a}, gives a positive $\Delta$R/R, where both eigenstates $\bra{\lambda_\pm}$ belong to the conduction band. In these intraband calculations, the eigenstates are accounted for by the availability of phonons up to $\sim$30 meV\cite{Gupta_2017}. It can be seen that the interband processes contribute higher to $\Delta$R/R as compared to the intraband processes, in agreement with the experimental observations where A$_1$ is more than A$_2$ for pressures below 9 GPa (Figure \ref{Fig:3p3}\textcolor{red}{a}). For comparing the experimental data, the values of calculated $\Delta$R/R were scaled to match the experimental values of A$_1$ (for interband) and A$_2$ (intraband) at ambient pressure, and the scaling factors were kept the same for all the pressures.
\par
In the context of the previously mentioned (pressure-independent) Hamiltonian, we derived a constant value for both interband and intraband contributions. This value effectively demonstrates a stable behavior within the pressure range of up to 3 GPa (Figure \ref{Fig:3p4}\textcolor{red}{d}). However, to comprehend the changes in $\Delta$R/R with pressure, it is imperative to tune the Hamiltonian in response to the applied pressure, namely opening of the bandgap in DSM Cd$_3$As$_2$ at a pressure 3 GPa\cite{Zhang2015}. Following the occurrence of the band gap opening at 3 GPa and subsequent increase, there is another change in the bandgap in the semiconducting phase at P=9 GPa (Figure \ref{Fig:3p4}\textcolor{red}{b}). The linear fits to the experimental data \cite{Zhang2015} (red curve in Figure \ref{Fig:3p4}\textcolor{red}{b}) is represented by $\eta'$, with the help of following equation:
\begin{align}
  \eta' &= 0\ \ \ \ \ \ \ \ \ \ \ \ \ \ \ \ for\ 0<P<3\ GPa \\
  &=(a_1+b_1P)\ \ \ \  for\ 3<P<9\ GPa \\
  &=(a_2+b_2P)\ \ \ \  for\ P>9\ GPa
\end{align}
where $a_1=-0.1117\ eV$, $b_1=3.026*10^{-2}\ eV/GPa$, $a_2=3.211*10^{-2}\ eV$, $b_2=3.684*10^{-2}\ eV/GPa$. For the pressures exceeding 3 GPa, the Hamiltonian is modified to $H_1=(\hbar^2/2m^*)(\sqrt{2}\eta k\sigma_x+\eta^2\sigma_y+k^2\sigma_z)$ where $\eta=\sqrt{\eta'm^*/\hbar^2}$, the effective mass, $m^*=0.04m_e$\cite{Moll_2016}, with $m_e$ representing the electronic mass. The energy eigenvalues of Hamiltonian $H_1$ are expressed as $\lambda_{1\pm}=\pm(\hbar^2/2m^*)(k^2+\eta^2)$ which depicts the dispersion relation for different pressure, illustrated in Figure \ref{Fig:3p4}\textcolor{red}{c}. The opening and gradual increase of the bandgap after P=3 GPa and a drastic change across P=9 GPa are well demonstrated. Figure \ref{Fig:3p4}\textcolor{red}{d} displays the experimental values of A$_1$ (depicted by blue circles) and A$_2$ (represented by red circles) alongside the corresponding calculated values (the black curve). The pressure-dependent values of A$_1$ and A$_2$ obtained from Hamiltonian $H_1$ show a close agreement between the calculated and experimental data. Consequently, attributing negative and positive components of $\Delta$R/R to interband (A$_1$) and intraband (A$_2$) contributions stands justified. 
\par
 The theoretical calculations demonstrated a notable alignment with the experimental value of A$_1$ and A$_2$ once we accounted for the band opening beyond the pressure of 3 GPa (Figure \ref{Fig:3p4}\textcolor{red}{d}). We have obtained the differential reflectivity at a particular time when it reaches its peak value at the electronic temperature of $T=652\ K$. Investigating the differential reflectivity at different electronic temperatures as the hot carriers cool down would be worth investigating.
\par
\begin{figure}[H]
    \centering
    \includegraphics[width=\textwidth]{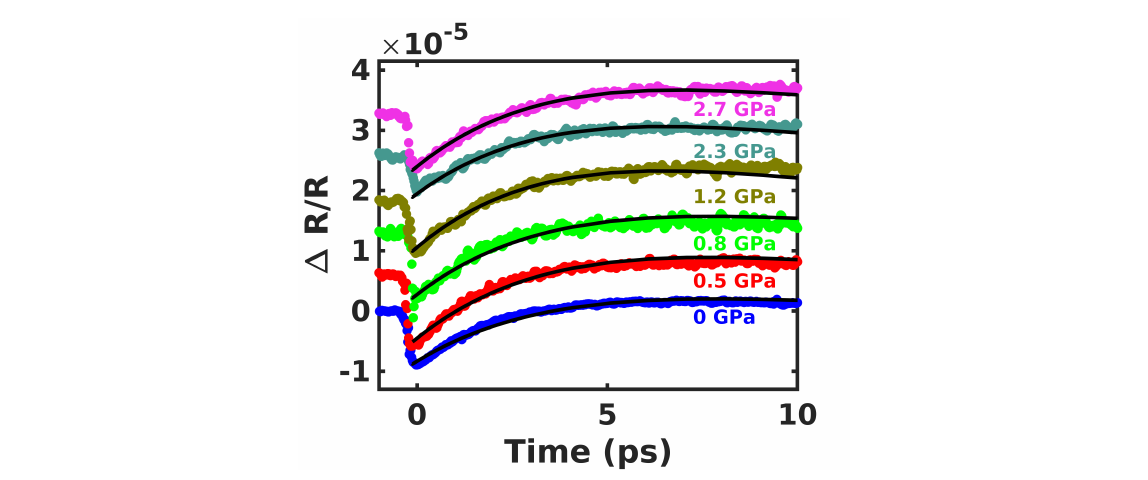}
    \caption[Two-temperature model elucidating the carrier dynamics in DSM Cd$_3$As$_2$]{The $\Delta R/R$ as a function of time for various pressures in the DSM phase of Cd$_3$As$_2$: The solid dots represent the data, and the black curves depict the reflectivity calculations as a function of electronic temperature, inherently related to time using the two-temperature model.}
    \label{Fig:3p5}
\end{figure}

We use the two-temperature model to describe the time evolution of electronic temperature ($T_e$) and lattice temperature ($T_L$) for a photoexcited state of a material\cite{Zhu2017,chen2006} as follows:
\begin{align}
     C_e\frac{dT_e}{dt}=-g_{e-ph}(T_e-T_L)+S(t)\ \ \ \ \ \ \ \ \ \ \ \ \ \\
    C_L\frac{dT_L}{dt}=g_{e-ph}(T_e-T_L)-B_{ph-ph}(T_L-T_0)
\end{align}
In this context, $C_e=70\ Jm^{-3}K^{-2}*T_e$ and $C_L=1.6*10^6\ Jm^{-3}K^{-1}$ represent the electronic and lattice-specific heat for Cd$_3$As$_2$ respectively\cite{Zhu2017}. The electron-phonon coupling, is denoted as $g_{e-ph}=5.3*10^{15}\ Wm^{-3}K^{-1}$, while the phonon-phonon coupling is represented by $B_{ph-ph}=9.4*10^9\ Wm^{-3}K^{-1}$\cite{Zhu2017}. The initial conditions, T$_e$(0)=T$_0$ and T$_L$(0)=T$_0$ (room temperature) have been chosen to solve the coupled equations. The source term S(t) is defined as,
\begin{equation}
    \centering
    S(t)=\sqrt{\frac{4\ln{2}}{\pi}}\frac{(1-R)F}{\delta_Pt_P}\exp\Bigg(-4\ln{2}\bigg(\frac{t-2t_P}{t_P}\bigg)^2\Bigg)
\end{equation}
The reflectivity is $R=0.46$ \cite{Katarzyna_1982}, the fluence is $F=243\ \mu J/cm^2$, the penetration depth is $\delta_P=125\ nm$ \cite{Zdanowicz_1967}, and the pulsewidth is $t_P=50\ fs$. The solution of these coupled equations provides us the electronic temperature as a function of time, $T_e(t)$, which is subsequently substituted in Eq.\ref{eq:3p3}-\ref{eq:3p6} to evaluate the differential reflectivity as a function of time, $\Delta R/R(t)$. Figure \ref{Fig:3p4}\textcolor{red}{d} illustrates that the theoretically calculated $A_1$ and $A_2$ (the black curve) have small deviations from the experimental values (blue and red circles). Therefore, after normalizing the calculated $\Delta R/R(0)$ to $A_1 + A_2$ for a given pressure, the time evolution of the calculated $\Delta R/R(t)$ aligns well with the experimental data, as shown in Figure \ref{Fig:3p5}. The solid colored circles represent the experimental data, whereas the black solid curves represent the calculated differential reflectivity with time. Thus, the theoretical calculations illustrate both the pressure-dependent trend of amplitudes $A_1$ and $A_2$ and the time evolution of the differential reflectivity.


\section*{Conclusions}
In summary, the optical pump optical probe spectroscopy of DSM Cd$_3$As$_2$ under high pressure is performed with an in-situ geometry. The fast amplitude and relaxation time with pressure demonstrates the transition of Cd$_3$As$_2$ across the pressures of 3 GPa and 9 GPa. The fast and slow relaxation processes have been attributed to interband and intraband transitions, respectively, illustrating the negative and positive contributions to the net differential reflectivity. The calculations captured the behavior of amplitudes $A_1$ and $A_2$ with pressure, necessitating the band opening beyond 3 GPa. This gap further increases with pressure. Using the two-temperature model, the electronic temperature as a function of time provided the time evolution of differential reflectivity, which closely matched the experimental data. A new rapid dynamical process is revealed beyond the pressure of 9 GPa which needs to be understood further. This study enhances our understanding of carrier relaxation dynamics under high pressure and highlights distinct phases of Cd$_3$As$_2$ at varying pressures, which could significantly influence the advancement of optoelectronic devices.

\section*{Acknowledgments} 
AKS thanks the Department of Science and Technology, Government of India, for financial support under the National Science Chair Professorship, and VA acknowledges CSIR for the research fellowship. VA thanks Mithun KP and Aindrila Sinha for insightful discussions. R.S. acknowledges the financial support provided by the Ministry of Science and Technology in Taiwan under Projects No. (NSTC 113–2124-M-001–003, 113–2112-M001-045-MY3, and 111-2124-M-A49-002).


\clearpage
\newpage
\bibliographystyle{unsrt}
\bibliography{citations_Cd3As2.bib}

\begin{thebibliography}{10}

\bibitem{Wang_2018}
Qi~Wang, Yuanfeng Xu, Rui Lou, Zhonghao Liu, Man Li, Yaobo Huang, Dawei Shen, Hongming Weng, Shancai Wang, and Hechang Lei.
\newblock {\em Nature Communications}, \textbf{9}(1):3681, 2018.

\bibitem{Shekhar_2015}
Chandra Shekhar, Ajaya~K. Nayak, Yan Sun, Marcus Schmidt, Michael Nicklas, Inge Leermakers, Uli Zeitler, Yurii Skourski, Jochen Wosnitza, Zhongkai Liu, Yulin Chen, Walter Schnelle, Horst Borrmann, Yuri Grin, Claudia Felser, and Binghai Yan.
\newblock {\em Nature Physics}, \textbf{11}(8):645--649, 2015.

\bibitem{Liang_2015}
Tian Liang, Quinnn Gibson, Mazhar~N. Ali, Minhao Liu, R.~J. Cava, and N.~P. Ong.
\newblock {\em Nature Materials}, \textbf{14}(3):280--284, 2015.

\bibitem{Higo_2018}
Tomoya Higo, Huiyuan Man, Daniel~B. Gopman, Liang Wu, Takashi Koretsune, Olaf M.~J. van~'t Erve, Yury~P. Kabanov, Dylan Rees, Yufan Li, Michi-To Suzuki, Shreyas Patankar, Muhammad Ikhlas, C.~L. Chien, Ryotaro Arita, Robert~D. Shull, Joseph Orenstein, and Satoru Nakatsuji.
\newblock {\em Nature Photonics}, \textbf{12}(2):73--78, 2018.

\bibitem{Ikhlas_2017}
Muhammad Ikhlas, Takahiro Tomita, Takashi Koretsune, Michi-To Suzuki, Daisuke Nishio-Hamane, Ryotaro Arita, Yoshichika Otani, and Satoru Nakatsuji.
\newblock {\em Nature Physics}, \textbf{13}(11):1085--1090, 2017.

\bibitem{Gooth_2018}
F.~Gooth, J.and~Menges, N.~Kumar, V.~S{\"u}$\beta$, C.~Shekhar, Y.~Sun, U.~Drechsler, R.~Zierold, C.~Felser, and B.~Gotsmann.
\newblock {\em Nature Communications}, \textbf{9}(1):4093, 2018.

\bibitem{Kumar_2021}
Nitesh Kumar, Satya~N. Guin, Kaustuv Manna, Chandra Shekhar, and Claudia Felser.
\newblock {\em Chemical Reviews}, \textbf{121}(5):2780--2815, 2021.

\bibitem{Young_2012}
S.~M. Young, S.~Zaheer, J.~C.~Y. Teo, C.~L. Kane, E.~J. Mele, and A.~M. Rappe.
\newblock {\em Phys. Rev. Lett.}, \textbf{108}:140405, 2012.

\bibitem{Yang_2014}
Bohm-Jung Yang and Naoto Nagaosa.
\newblock {\em Nature Communications}, \textbf{5}(1):4898, 2014.

\bibitem{Liu_2014}
J.~Liu, Z. K.and~Jiang, B.~Zhou, Z.~J. Wang, Y.~Zhang, H.~M. Weng, D.~Prabhakaran, S.-K. Mo, H.~Peng, P.~Dudin, T.~Kim, M.~Hoesch, Z.~Fang, X.~Dai, Z.~X. Shen, D.~L. Feng, Z.~Hussain, and Y.~L. Chen.
\newblock {\em Nature Materials}, \textbf{13}(7):677--681, 2014.

\bibitem{Burkov_2011}
A.~A. Burkov, M.~D. Hook, and Leon Balents.
\newblock {\em Phys. Rev. B}, \textbf{84}:235126, 2011.

\bibitem{Wang_2013}
Zhijun Wang, Hongming Weng, Quansheng Wu, Xi~Dai, and Zhong Fang.
\newblock {\em Phys. Rev. B}, \textbf{88}:125427, 2013.

\bibitem{Jeon_2014}
Sangjun Jeon, Brian~B. Zhou, Andras Gyenis, Benjamin~E. Feldman, Itamar Kimchi, Andrew~C. Potter, Quinn~D. Gibson, Robert~J. Cava, Ashvin Vishwanath, and Ali Yazdani.
\newblock {\em Nature Materials}, \textbf{13}(9):851--856, 2014.

\bibitem{Yi_2014}
Hemian Yi, Zhijun Wang, Chaoyu Chen, Youguo Shi, Ya~Feng, Aiji Liang, Zhuojin Xie, Shaolong He, Junfeng He, Yingying Peng, Xu~Liu, Yan Liu, Lin Zhao, Guodong Liu, Xiaoli Dong, Jun Zhang, M.~Nakatake, M.~Arita, K.~Shimada, H.~Namatame, M.~Taniguchi, Zuyan Xu, Chuangtian Chen, Xi~Dai, Zhong Fang, and X.~J. Zhou.
\newblock {\em Scientific Reports}, \textbf{4}(1):6106, 2014.

\bibitem{Neupane_2014}
Madhab Neupane, Su-Yang Xu, Raman Sankar, Nasser Alidoust, Guang Bian, Chang Liu, Ilya Belopolski, Tay-Rong Chang, Horng-Tay Jeng, Hsin Lin, Arun Bansil, Fangcheng Chou, and M.~Zahid Hasan.
\newblock {\em Nature Communications}, \textbf{5}(1):3786, 2014.

\bibitem{Borisenko_2014}
Sergey Borisenko, Quinn Gibson, Danil Evtushinsky, Volodymyr Zabolotnyy, Bernd B\"uchner, and Robert~J. Cava.
\newblock {\em Phys. Rev. Lett.}, \textbf{113}:027603, 2014.

\bibitem{Rosenberg_1959}
Arthur~J. Rosenberg and Theodore~C. Harman.
\newblock {\em Journal of Applied Physics}, \textbf{30}:1621, 1959.

\bibitem{Li_2015}
Cai-Zhen Li, Li-Xian Wang, Haiwen Liu, Jian Wang, Zhi-Min Liao, and Da-Peng Yu.
\newblock {\em Nature Communications}, \textbf{6}(1):10137, 2015.

\bibitem{Narayanan_2015}
A.~Narayanan, M.~D. Watson, S.~F. Blake, N.~Bruyant, L.~Drigo, Y.~L. Chen, D.~Prabhakaran, B.~Yan, C.~Felser, T.~Kong, P.~C. Canfield, and A.~I. Coldea.
\newblock {\em Phys. Rev. Lett.}, \textbf{114}:117201, 2015.

\bibitem{Sun2016}
Zhipei Sun, Amos Martinez, and Feng Wang.
\newblock {\em Nature Photonics}, \textbf{10}(4):227--238, 2016.

\bibitem{Xia2009}
Fengnian Xia, Thomas Mueller, Yu-ming Lin, Alberto Valdes-Garcia, and Phaedon Avouris.
\newblock {\em Nature Nanotechnology}, \textbf{4}(12):839--843, 2009.

\bibitem{Martinez2013}
Amos Martinez and Zhipei Sun.
\newblock {\em Nature Photonics}, \textbf{7}(11):842--845, 2013.

\bibitem{Yao_2021}
Xiaomei Yao, Shengxi Zhang, Qiang Sun, Peizong Chen, Xutao Zhang, Libo Zhang, Jian Zhang, Yan Wu, Jin Zou, Pingping Chen, and Lin Wang.
\newblock {\em ACS Photonics}, \textbf{8}(6):1689--1697, 2021.

\bibitem{Wang_2012}
Zhijun Wang, Yan Sun, Xing-Qiu Chen, Cesare Franchini, Gang Xu, Hongming Weng, Xi~Dai, and Zhong Fang.
\newblock {\em Phys. Rev. B}, \textbf{85}:195320, 2012.

\bibitem{Liu2014}
Z.~K. Liu, B.~Zhou, Y.~Zhang, Z.~J. Wang, H.~M. Weng, D.~Prabhakaran, S.-K. Mo, Z.~X. Shen, Z.~Fang, X.~Dai, Z.~Hussain, and Y.~L. Chen.
\newblock {\em Science}, \textbf{343}(6173):864--867, 2014.

\bibitem{Wang2017}
Qinsheng Wang, Cai-Zhen Li, Shaofeng Ge, Jin-Guang Li, Wei Lu, Jiawei Lai, Xuefeng Liu, Junchao Ma, Da-Peng Yu, Zhi-Min Liao, and Dong Sun.
\newblock {\em Nano Letters}, \textbf{17}(2):834--841, 2017.

\bibitem{Meng_2018}
Yafei Meng, Chunhui Zhu, Yao Li, Xiang Yuan, Faxian Xiu, Yi~Shi, Yongbing Xu, and Fengqiu Wang.
\newblock {\em Opt. Lett.}, \textbf{43}(7):1503--1506, 2018.

\bibitem{WeiLu_2018}
Wei Lu, Jiwei Ling, Faxian Xiu, and Dong Sun.
\newblock {\em Phys. Rev. B}, \textbf{98}:104310, 2018.

\bibitem{Weber_2015}
C.~P. Weber, Ernest Arushanov, Bryan~S. Berggren, Tahereh Hosseini, Nikolai Kouklin, and Alex Nateprov.
\newblock {\em Applied Physics Letters}, \textbf{106}(23), 2015.

\bibitem{WeiLu_2017}
Wei Lu, Shaofeng Ge, Xuefeng Liu, Hong Lu, Caizhen Li, Jiawei Lai, Chuan Zhao, Zhimin Liao, Shuang Jia, and Dong Sun.
\newblock {\em Phys. Rev. B}, \textbf{95}:024303, 2017.

\bibitem{Zhu2017}
Chunhui Zhu, Xiang Yuan, Faxian Xiu, Chao Zhang, Yongbing Xu, Rong Zhang, Yi~Shi, and Fengqiu Wang.
\newblock {\em Applied Physics Letters}, \textbf{111}(9), 2017.

\bibitem{Zhu_2017}
Chunhui Zhu, Fengqiu Wang, Yafei Meng, Xiang Yuan, Faxian Xiu, Hongyu Luo, Yazhou Wang, Jianfeng Li, Xinjie Lv, Liang He, Yongbing Xu, Junfeng Liu, Chao Zhang, Yi~Shi, Rong Zhang, and Shining Zhu.
\newblock {\em Nature Communications}, \textbf{8}(1):14111, 2017.

\bibitem{Liang_2017}
Tian Liang, Satya Kushwaha, Jinwoong Kim, Quinn Gibson, Jingjing Lin, Nicholas Kioussis, Robert~J. Cava, and N.~Phuan Ong.
\newblock {\em Science Advances}, \textbf{3}(5):e1602510, 2017.

\bibitem{Qi_2016}
Pavel~G. Qi, Yanpengand~Naumov, Catherine~R. Ali, Mazhar N.and~Rajamathi, Walter Schnelle, Oleg Barkalov, Michael Hanfland, Shu-Chun Wu, Chandra Shekhar, Yan Sun, Vicky S{\"u}{\ss}, Marcus Schmidt, Ulrich Schwarz, Eckhard Pippel, Peter Werner, Reinald Hillebrand, Tobias F{\"o}rster, Erik Kampert, Stuart Parkin, R.~J. Cava, Claudia Felser, Binghai Yan, and Sergey~A. Medvedev.
\newblock {\em Nature Communications}, \textbf{7}(1):11038, 2016.

\bibitem{Dissanayake_2019}
Sachith Dissanayake, Chunruo Duan, Junjie Yang, Jun Liu, Masaaki Matsuda, Changming Yue, John~A. Schneeloch, Jeffrey C.~Y. Teo, and Despina Louca.
\newblock {\em npj Quantum Materials}, \textbf{4}(1):45, 2019.

\bibitem{Reis_2016}
R.~D. dos Reis, S.~C. Wu, Y.~Sun, M.~O. Ajeesh, C.~Shekhar, M.~Schmidt, C.~Felser, B.~Yan, and M.~Nicklas.
\newblock {\em Phys. Rev. B}, \textbf{93}:205102, 2016.

\bibitem{Braun_2018}
Johannes~M Braun, Harald Schneider, Manfred Helm, Rafał Mirek, Lynn~A Boatner, Robert~E Marvel, Richard~F Haglund, and Alexej Pashkin.
\newblock {\em New Journal of Physics}, \textbf{20}(8):083003, 2018.

\bibitem{Wu_2024}
Simin Wu, Weibin Chu, Yang Lu, and Minbiao Ji.
\newblock {\em Nano Letters}, \textbf{24}(1):424--432, 2024.
\newblock PMID: 38153402.

\bibitem{Zhang_2024}
Kai Zhang, Jiafeng Xie, Jin Yang, Tianwu Wang, Fuhai Su, Yirong Wu, and Guangyou Fang.
\newblock {\em Phys. Rev. B}, \textbf{109}:115130, 2024.

\bibitem{Tu_2023}
Hongyu Tu, Lingyun Pan, Hongjian Qi, Shuhao Zhang, Fangfei Li, Chenglin Sun, Xin Wang, and Tian Cui.
\newblock {\em J Phys Condens Matter}, \textbf{35}(25), 2023.

\bibitem{Fotev_2023}
Ivan Fotev, Stephan Winnerl, Saicharan Aswartham, Sabine Wurmehl, Bernd B\"uchner, Harald Schneider, Manfred Helm, and Alexej Pashkin.
\newblock {\em Phys. Rev. B}, \textbf{108}:035101, 2023.

\bibitem{Yang_2024}
Y.~Yang, Y.~H. Meng, B.~R. Lu, F.~Jin, Y.~G. Shi, F.~Hong, S.~S. Zhang, X.~H. Yu, X.~B. Wang, and J.~L. Luo.
\newblock {\em Phys. Rev. B}, \textbf{109}:064307, 2024.

\bibitem{Zhang2015}
Shan Zhang, Qi~Wu, Leslie Schoop, Mazhar~N. Ali, Youguo Shi, Ni~Ni, Quinn Gibson, Shang Jiang, Vladimir Sidorov, Wei Yi, Jing Guo, Yazhou Zhou, Desheng Wu, Peiwen Gao, Dachun Gu, Chao Zhang, Sheng Jiang, Ke~Yang, Aiguo Li, Yanchun Li, Xiaodong Li, Jing Liu, Xi~Dai, Zhong Fang, Robert~J. Cava, Liling Sun, and Zhongxian Zhao.
\newblock {\em Phys. Rev. B}, \textbf{91}:165133, 2015.

\bibitem{Gupta_2017}
Satyendra~Nath Gupta, D.~V.~S. Muthu, C.~Shekhar, R.~Sankar, C.~Felser, and A.~K. Sood.
\newblock {\em Europhysics Letters}, \textbf{120}(5):57003, 2018.

\bibitem{Mao1978}
H.~K. Mao, P.~M. Bell, J.~W. Shaner, and D.~J. Steinberg.
\newblock {\em Journal of Applied Physics}, \textbf{49}(6):3276--3283, 1978.

\bibitem{Mao1986}
H.~K. Mao, J.~Xu, and P.~M. Bell.
\newblock {\em Journal of Geophysical Research: Solid Earth}, \textbf{91}(B5):4673--4676, 1986.

\bibitem{Prut2022}
V.~V. Prut.
\newblock {\em Russian Physics Journal}, \textbf{65}(7):1172--1178, 2022.

\bibitem{Sankar_2015}
R~Sankar, M~Neupane, S-Y Xu, C~J Butler, I~Zeljkovic, I~Panneer~Muthuselvam, F-T Huang, and S-T Guo.
\newblock \textbf{5}:12966, 2015.

\bibitem{Takeda_2024}
K.~S. Takeda, K.~Uchida, K.~Nagai, S.~Kusaba, S.~Takahashi, and K.~Tanaka.
\newblock {\em Phys. Rev. Lett.}, \textbf{132}:186901, 2024.

\bibitem{chen2006}
JK~Chen, DY~Tzou, and JE~Beraun.
\newblock {\em International journal of heat and mass transfer}, \textbf{49}(1-2):307--316, 2006.

\bibitem{Wang2019}
Xiaofan Wang, Keisuke Shinokita, Hong~En Lim, Nur~Baizura Mohamed, Yuhei Miyauchi, Nguyen~Thanh Cuong, Susumu Okada, and Kazunari Matsuda.
\newblock {\em Advanced Functional Materials}, \textbf{29}(6):1806169, 2019.

\bibitem{Conte2017}
Adriano Mosca~Conte, Olivia Pulci, and Friedhelm Bechstedt.
\newblock {\em Scientific Reports}, \textbf{7}(1):45500, 2017.

\bibitem{Neto_2009}
A.~H. Castro~Neto, F.~Guinea, N.~M.~R. Peres, K.~S. Novoselov, and A.~K. Geim.
\newblock {\em Rev. Mod. Phys.}, \textbf{81}:109--162, 2009.

\bibitem{Moll_2016}
Philip J.~W. Moll, Nityan~L. Nair, Toni Helm, Andrew~C. Potter, Itamar Kimchi, Ashvin Vishwanath, and James~G. Analytis.
\newblock {\em Nature}, \textbf{535}(7611):266--270, 2016.

\bibitem{Katarzyna_1982}
Katarzyna Karnicka-Moscicka, Andrzej Kisiel, and Lidia Zdanowicz.
\newblock {\em Solid State Communications}, \textbf{44}(3):373--377, 1982.

\bibitem{Zdanowicz_1967}
L.~Żdanowicz.
\newblock {\em physica status solidi (b)}, \textbf{20}(2):473--480, 1967.

\end{thebibliography}

\end{document}